\documentclass[11pt, a4paper]{article}
\usepackage{jheppub,dsfont,epsfig}

\newcommand{\eq}[1]{eq.~(\ref{#1})}

\newcommand{\Eq}[1]{Eq.~(\ref{#1})}

\newcommand{\ur}[1]{(\ref{#1})}

\newcommand{\beq}{\begin{equation}}
\newcommand{\eeq}{\end{equation}}
\newcommand{\la}[1]{\label{#1}}
\newcommand{\bea}{\begin{eqnarray}}
\newcommand{\eea}{\end{eqnarray}}
\newcommand{\ba}{\begin{array}}
\newcommand{\ea}{\end{array}}

\newcommand{\half}{{\textstyle{\frac{1}{2}}}}

\newcommand{\Tr}{{\rm Tr}}

\title{Effective Lagrangian for the Polyakov line on a lattice\footnote{On December 26 2012
 our esteemed colleague Dmitri Diakonov passed away untimely early.
 At that time he was working with one of us (V.P.) on a follow up
 project to a paper we had published in June of that year. This article
 is based on this cooperation and we dedicate it to the memory of our friend
 Dmitri Diakonov.}}
\author[a, b]{Dmitri Diakonov,}
\author[a]{Victor Petrov,}
\author[c]{Hans-Peter Schadler$\,$}
\author[c]{and Christof Gattringer$\,$}

\affiliation[a]{Petersburg Nuclear Physics Institute, Kurchatov National Research Centre\\
Gatchina, St. Petersburg 188300, Russia}

\affiliation[b]{St. Petersburg Academic University, St. Petersburg 194021, Russia}

\affiliation[c]{Institute for Physics, Karl-Franzens-Universit\"at Graz,
Universit\"atsplatz 5, 8010 Graz, Austria \vskip5mm}

\emailAdd{victorp@thd.pnpi.spb.ru}
\vspace{1mm}
\emailAdd{hps@abyle.org}
\vspace{1mm}
\emailAdd{christof.gattringer@uni-graz.at}

\abstract{We formulate a method for computing the effective Lagrangian of the Polyakov line
on the lattice. Using mean field approximation we calculate the effective potential for
high temperatures. The result agrees with recent lattice simulations. We reveal a new
type   of ultraviolet divergence (coming from longitudinal gluons) which dominates the
effective potential and explains the discrepancy of the  lattice simulations and standard
perturbative calculations performed in covariant gauges.}

\arxivnumber{}

\begin{document}
\maketitle
\flushbottom

\section{Introduction}

The Polyakov line, defined as the path-ordered exponential of the time component of the
Yang--Mills connection in Euclidian signature,
\beq
{\mathcal P}(\vec{x})=P\exp\left(i\int_0^{1/T}\! dt\, A_4(\vec{x},t)\right)\,,
\la{Polyakov-def}\eeq
where $T$ is the temperature, is an order parameter for the confinement-deconfinement
phase transition in pure gauge theory~\cite{Polyakov-line}. Its average over the
gauge-field ensemble behaves as
\beq
\langle\,\Tr{\mathcal P}(\vec{x})\,\rangle\quad
\left\{\begin{array}{ccc}
= 0 & {\rm at}\;T<T_c\,,\;\; & {\rm confinement},\\
\neq 0 & {\rm at}\;T>T_c\,,\;\; &{\rm deconfinement}.\\
\end{array}
\right.
\la{order}\eeq
It would be helpful to know the precise effective Lagrangian for this important variable in order
to understand better what physical mechanisms govern the deconfinement phase transition.

Quite recently the effective potential, {\it i.e.}, the derivative-independent part of the
effective Lagrangian has been found for the $SU(2)$ and $SU(3)$ gauge groups by direct
numerical simulations on the lattice~\cite{DGS}, with puzzling results: At high
temperatures ($T\leq 5T_c$) the effective potential as function of the eigenvalues of
${\cal P}$ did not follow the expected long-known perturbative potential~\cite{GPY,W,DO}
but turned out to be a factor of 20 to 30 times larger, and only suitable ratios of
observables could be matched to the perturbative results.

In this paper we first define accurately the effective Lagrangian for the Polyakov loop
by using the lattice regularization in the ultraviolet. This is also necessary for a
direct comparison with the lattice results~\cite{DGS}. Second, we develop a systematic
mean-field method for computing the lattice-regularized effective Lagrangian. The method
is in fact a well-defined loop expansion about the mean field with the expansion
parameter being $1/\beta$ where $\beta$ is the lattice inverse gauge coupling. Special
care is taken in dealing with the zero modes in that expansion. Third, we compare the
results of our analytical calculations with the numerical lattice results~\cite{DGS}. We
find good agreement between the two already at the one loop level. We thus resolve the
above mentioned puzzle in the numerical results and discuss the reason for the large
discrepancy between lattice results and the perturbative potential~\cite{GPY,W,DO}.
Fourth, our expansion about the mean field solution paves the way for computing terms
with the spatial derivatives of the Polyakov loop eigenvalues in the effective
Lagrangian, which can be also found on the lattice.

\section{Lattice partition function as a path integral over the eigenvalues of the Polyakov line}

The Yang--Mills partition function defined as a path integral over the connection
$A_\mu(\vec{x},t)$ with periodic boundary conditions can be identically rewritten in such
a way that the integration over ${\cal P}(\vec{x}\,)$ is performed last:
\beq
{\cal Z_{\rm contin}}=\int\!D [A] \,\exp\left(-\frac{1}{2g^2}\int\!d^{\,4} x\,\Tr F_{\mu\nu}F_{\mu\nu}\right)
=\int\!D[{\cal P}]\,\exp\left(-S_{\rm eff}[{\cal P}]\right)\,.
\la{Seff-def-1}\eeq
This equation formally defines what we call the effective action $S_{\rm eff}$ and the
effective Lagrangian ${\cal L}_{\rm eff}$ for the Polyakov line, $S_{\rm eff}[{\cal
P}]=\int\!d^{\,3} x\,{\cal L}_{\rm eff}[{\cal P}]$.

Let us rewrite \Eq{Seff-def-1} using the standard lattice regularization of pure
Yang--Mills theory. The partition function is defined as the path integral over link
variables $U_\mu(x) \in SU(N)$,
\beq
{\cal Z}=\int D[U] e^{-\beta S[U]} \; ,
\la{def-1}
\eeq
where $S[U]$ is the standard Wilson plaquette action, and $\beta = \frac{2N}{g^2}$  is
the inverse bare gauge coupling constant. The hypercubic lattice has $N_s$ sites in the
spatial directions and  $N_t$ sites in the Euclidean time direction. The corresponding
coordinates assume values $x_1, x_2, x_3 = 0,1, \, ... \, N_s-1$ and $x_4 = t = 0,1, \,
... \, N_t-1$. We use $x$ to denote the full vector with all 4 components, $t$ is used
for the time component (= 4-component) and $\vec{x}$ for the spatial part. The lattice
spacing is mostly set to $a = 1$ in this paper and  we display $a$ only where necessary.
The link variables $U_\mu(x)$ are assumed to satisfy periodic  boundary conditions in all
four directions. The integration measure $D[U]$ is the product of invariant Haar
measures  for all $U_\mu(x)$ normalized to unity. We will use the standard Wilson action
\beq
S[U]=\sum_{x, \mu < \nu}\left[1-\frac{1}{N}{\rm Re}\Tr U_{\mu\nu}(x)\right], \qquad
U_{\mu\nu}(x)=U_\mu(x)U_\nu(x+\hat{\mu})U_\mu^\dagger(x+\hat{\nu}) U_\nu^\dagger(x),
\la{Wilson}
\eeq
where the sum runs over all plaquettes. The temperature is given by $T=\Lambda/N_t$ where
$\Lambda=\frac{1}{a}\,f(\beta)$ is a renormalization-group-invariant combination. Usually
one uses the relative temperature, say, the ratio of the temperature to the temperature
of the phase transition $T_c$ which should be determined separately for a given setup.
Then $\frac{T}{T_c}=\frac{N_{t}^{(c)}}{N_t}$, where $N_{t}^{(c)}$ is the critical
temporal extent of the lattice.

The partition function \ur{def-1} implies the integration over all possible gauges and
therefore contains the volume of the gauge group. The volume, however, is unity according
to our definition of the Haar measure. Let us rewrite \eq{def-1} in the form which is
close to the physical gauge $A_4=0$ in the continuum. Using standard arguments (see,
e.g., \cite{GaLa}) one can fix the gauge links on a tree (a structure of links  on the
lattice without closed loops) to arbitrary values. This allows us to fix all temporal
gauge links to $U_4(\vec{x},t) = \mathds{1}$ with $t = 0,1, \, ... N_t -2$, except for
the last time-slice, i.e., the temporal links $U_4(\vec{x},N_t-1)$. The Polyakov line at
$\vec{x}$  then reduces to the temporal link at this last timeslice
\[
\mathcal{P}(\vec{x})\; \equiv \; \prod_{t=0}^{N_t-1}
U_4(\vec{x},t) \; = \; U_4(\vec{x},N_t-1) \; .
\]
We arrive at the following representation for the statistical sum:
\beq
{\cal Z} \; = \; \int D[{\mathcal P}] \, \int D[U_j] \;
e^{-\beta {\mathcal S}[U_j,{\mathcal P}]} \; .
\la{def-12}
\eeq
The action $\mathcal {S}[U_j,{\mathcal P} ]$ is the Wilson action in the
gauge where all temporal links are trivial except those on the last time slice
where they are given by ${\mathcal P}(\vec{x})$.

The Polyakov lines ${\mathcal P}(\vec{x}) \in SU(N)$ are diagonalized by
matrices $Q(\vec{x})$ such that
\beq
Q(\vec{x}) \, {\mathcal P}(\vec{x}) \, Q(\vec{x})^\dagger \; = \;
{\rm diag} \left(e^{i\varphi_1(\vec{x})},\ldots, e^{i\varphi_r(\vec{x})},
e^{-i\sum_{n=1}^r \varphi_n(\vec{x})} \right),
\eeq
where $r=N-1$ is the rank of the group. With a gauge transformation independent of time,
which leaves the already trivial temporal links unchanged, diagonalizes the ${\mathcal
P}(\vec{x})$, and transforms the spatial link variables as
\beq
U_j(\vec{x},t)\; \to \; Q(\vec{x})\, U_j(\vec{x},t) \, Q(\vec{x}+\widehat{j}\,)^\dagger \; ,
\eeq
one can reduce the matrices ${\mathcal P}(\vec{x})$ to diagonal form in the partition
sum, i.e., \eq{def-12} depends only on the eigenvalues of the Polyakov lines.  Hence, the
integration measure $D[{\mathcal P}]$ in \eq{def-12} can be  substituted by the
integration over phases $\varphi_1(\vec{x})\ldots \varphi_r(\vec{x})$  with the
corresponding measure. Note that the eigenvalues of the Polyakov lines ${\mathcal
P}(\vec{x})$  (i.e., the phases $\varphi_i(\vec{x}$)) are {\em gauge invariant}.

We thus define the effective action for the Polyakov line on the lattice as:
\beq
e^{-\beta \, S_{\rm eff}({\mathcal P})}=\int D[U_j]  \;e^{-\beta \, {\mathcal S}[U_j,{\mathcal P} ]} \; ,
\la{def-2}
\eeq
which is a path integral over only the spatial links $U_j$.  The effective action defined
in \eq{def-2} depends only on the eigenvalues of the  matrices ${\mathcal P}(\vec{x})$.
The full partition sum and moments of Polyakov lines  can be found by integrating this
action over the $\varphi_n(\vec{x})$ (including the necessary measure factors).

Yet another gauge transformation $U_\mu(x) \to S(x)\, U_\mu(x) \,S(x +
\widehat{\mu})^\dagger$ with
\beq
S(\vec{x},t)={\rm diag}\left(e^{i\frac{N_t-t}{N_t}\varphi_1(\vec{x})},
\ldots,
e^{-i\frac{N_t-t}{N_t}\sum_{n = 1}^r\varphi_n(\vec{x})}\right),
\qquad t=0,\ldots N_t - 1 \; ,
\eeq
is used to make all temporal gauge links independent of time,
\beq
U_4(\vec{x},t)={\rm diag}\left(e^{i\frac{\varphi_1(\vec{x})}{N_t}},
\ldots e^{-i \frac{1}{N_t}\sum_{n=1}^r \varphi_n(\vec{x})}\right),\quad t = 0,N_t-1 \; .
\label{u4final}
\eeq
This gauge is the one most suitable for our purposes.

In this paper we will consider the effective action only for phases $\varphi_n$
that are constant in space. In other words,  we are going
to calculate an {\em effective potential} for spatially constant Polyakov lines.
We will restrict ourselves to the case of gauge group $SU(2)$.
It is straightforward to generalize these calculations to other gauge groups.

Numerical studies of the effective action ${\cal S}_{\rm eff}[{\mathcal P}]$
for spatially constant $\varphi_n$ were presented in Ref.~\cite{DGS}
for both gauge groups $SU(2)$ and $SU(3)$. Recent results for the effective
potential from other approaches
can be found in \cite{others}.

Another study of an effective potential was presented recently in 
Ref.~\cite{SDPS}, where the {\em per-site} potential for the Polyakov line was 
calculated. This object is different
from the effective potential discussed here and aims at analyzing gradient 
terms. In principle, the per-site quantity can be found in our setting by
integrating \eq{def-2} over all Polyakov lines ${\cal P}({\bf x})$ 
except for one spatial site. 


\section{Mean field approximation}

To compute the statistical sum ${\mathcal Z}$ and the effective action  ${\mathcal
S}_{\rm eff}[{\mathcal P}]$ one can use mean field approximation  which is known to be
applicable at sufficiently large $\beta$. Of course, mean  field approximation is not
able to describe confinement -- this approximation is  nothing more than a modification
of perturbation theory. Nevertheless, it can provide  an estimate for the effective
action valid at large $\beta$ and/or at large temperatures.

For our mean field calculation we will follow the approach of \cite{Druffe} where mean
field  theory for lattice fields theories was considered as a version of saddle point
approximation.

We identically rewrite the expression \eq{def-2} for the effective action as:
\beq
e^{-\beta{\cal S}_{\rm eff}[{\mathcal P}]}=\int\! \frac{D[H_j] \, D[V_j]}{2\pi} \int \! D[U_j] \, \exp \!
\left(-\beta \, {\mathcal S}[V_j,{\mathcal P}]+{\rm Re}\sum_{x,j}
\Tr[H^\dagger_j(x) (U_j(x) - V_j(x))]
\right) .
\eeq
Here $H_j$ and $V_j$ are suitable $N\times N$ matrices defined on the spatial links,
which we will specify  in more detail below. The $H_j$ are chosen such that integration
$\int D[H_j]$ produces the  $\delta$-function $\delta(U-V)$. Subsequent integration over
the $V_j$ returns one to the original expression (\ref{def-2}).

Let us now first integrate out the original spatial link variables.  This integration
leads to local functionals $W[H_j(x)]$ for each spatial link,
\beq
e^{W[H_j(x)]} \; = \; \int d U_j(x) \exp\left( {\rm Re}\Tr[H^\dagger_j(x) U_j(x)]\right) \; ,
\eeq
which can be calculated explicitly for the given gauge group ($dU_j(x)$  denotes Haar
measure integration for a single element of the gauge group). The remaining integral,
\beq
e^{-\beta{\cal S}_{\rm eff}[{\mathcal P}]} \; = \; \int \frac{D[H_j] D[V_j]}{2\pi} \, \exp
\left(-\beta \, {\mathcal S}[V_j,{\mathcal P}] \, +\, \sum_{x,j} \left[
W[H_j(x)]-{\rm Re}\Tr[H^\dagger_j(x) V_j(x)] \right]
\right) \; ,
\la{def-4}
\eeq
can be calculated in the saddle point approximation. The saddle point $\overline{H}_j$,
$\overline{V}_j$ can be found from the ''equations of motion'',
\beq
\frac{\delta{\mathcal S}}{\delta V_j(x)} \, = \, H^\dagger_j(x),
\qquad \frac{\delta W}{\delta H_j(x)} \, = \, V_j(x) \; .
\la{eq-m}
\eeq
We are looking for the translationally invariant and isotropic solution proportional to
the  unit matrix:
\beq
\overline{H}_j(x) \; = \; h\cdot \mathds{1} \; , \qquad \overline{V}_j(x) = v \cdot \mathds{1} \; .
\la{s-p}
\eeq
For such configurations the Wilson action \eq{Wilson} reduces to
\beq
\frac{S[\overline{V}_j,{\mathcal P}]}{N^3_sN_t} \; = \;
\frac{d(d-1)}{2}-(d-1)\left[v^2+\frac{(d-2)}{2}v^4\right] \; ,
\eeq
where $N^3_sN_t$ is the total number of lattice sites and $d=4$ is the  dimension of
Euclidean space-time. Let us note that for the case of  constant phases $\varphi_n$ the
dependence on the  Polyakov line disappears from the saddle point action. It will
re-appear only in the 1-loop correction.

The calculation of $W[H_j(x)]$ is possible for any gauge group.  However, here we will
restrict ourselves to the group $SU(2)$ to keep things as simple  as possible. In this
case we need only one phase in (\ref{u4final}) which we denote as $\varphi$. The group
$SU(2)$ has some specifics \cite{Druffe}: The matrices $H_j(x)$ and $V_j(x)$ are
Hermitian,  which is related to the fact that $SU(2)$ is a self-conjugated group.  We
parameterize them as ($\alpha = 1,2,3,4$, summed),
\beq
U=u_\alpha \sigma_\alpha^{-}, \qquad V=v_\alpha\sigma^-_\alpha,
\qquad H=h_\alpha\sigma^-_\alpha, \qquad \sigma^\pm_\alpha = (\mp i\vec{\sigma},
\mathds{1}) \; .
\eeq
For the unitary matrix $U$ we have the additional constraint $u_\alpha^2 = 1$.
The integral determining $W[H_j(x)] = W[h]$ becomes:
\beq
e^{W[h]}=\int\frac{d^4u}{2\pi^2} \;
\delta(u^2_\alpha-1)\,e^{h u_4 } \;\;, \;\; \;  \mbox{such that} \;\;  W[h]=\log\frac{2I_1(h)}{h} \; .
\eeq
Here $2\pi^2$ is the volume of $SU(2)$, and $I_1$ a modified Bessel function.
The equations of motion (\ref{eq-m}) assume the form
\beq
h=2\beta(v+2v^3), \qquad  v=\frac{I_2(h)}{I_1(h)} \; .
\la{eqm1}
\eeq
They have non-trivial solutions for $\beta>\beta_c\approx 1.6817$. At large $\beta\gg 1$
the mean field $v$ tends towards $1$ and $h$ becomes large. The statistical sum in the
saddle point approximation is given by
\beq
\log{\mathcal Z}_0= 3N^3_sN_t\left[
\log\left(\frac{2I_1(h)}{h}\right)-\beta v^2(1+3v^2)-6\beta
\right] \; .
\la{mf}
\eeq
At the critical point $\beta_c$ the mean field approximation predicts a phase
transition.  For gauge group $SU(2)$ this transition is fictitious -- it appears due to
the fact that our  approximation is rather crude. However, at larger $\beta$ mean field
approximation  describes the lattice data (e.g., for the average plaquette energy) quite
accurately  (see, e.g., \cite{Druffe}).

\section{Loop corrections to the mean field solution}

As already remarked, in the mean field approximation the free energy of  $SU(2)$ lattice
gauge theory does depend neither on temperature nor on the Polyakov line. One needs to
compute 1-loop corrections in order to study this dependence.

To obtain the first correction to the saddle point approximation we consider quantum
fluctuations for $V_j$ and $H_j$ around the saddle points and parameterize $V_j$ and
$H_j$ entering \eq{def-4} as ($\alpha = 1,2,3,4$ is summed,
$\sigma^\pm_\alpha = (\mp i\vec{\sigma}, \mathds{1})$ ):
\beq
V_j(x)=v\cdot \mathds{1}+w^\alpha_j(x) \sigma^-_\alpha \; ,
\qquad H_j(x)=h\cdot\mathds{1}+\rho^\alpha_j(x)
\sigma^-_\alpha \; ,
\eeq
and in the path integrals the measures $D[V_j]$ and $D[H_j]$ are replaced by the
corresponding  measures for the parameters $w^\alpha_j(x)$, $h^\alpha_j(x)$, which we
denote as $D[w_j]$ and $D[\rho_j]$. We assume that the quantum fluctuations
$w^\alpha_j(x)$ and $h^\alpha_j(x)$ are small and  expand the action in these quantum
fluctuations. The linear term disappears due to the equations of motion. For computing
the quadratic terms  we begin with the corrections to $W[H_j(x)]$:
\beq
W[H_j(x)]=W_{0}[h]+\frac{I_2(h)}{2hI_1(h)}\left[\frac{(\rho^4_j(x))^2}{\kappa^2}+ (\vec{\rho}_j(x))^{\;2}
\right] \, \; ,
\; \;
\kappa^{-2}=1+\frac{hI_3(h)}{I_2(h)}-\frac{hI_2(h)}{I_1(h)} \; ,
\eeq
where $W_0[h]$ is the saddle point value. It can be seen that
$\kappa$ is always larger than unity and monotonically increases with increasing $h$.

The integrals over the $\rho_j^\alpha (x)$ are Gaussian
and can be performed easily, such that we end up with:
\begin{eqnarray}
e^{-\beta{\cal S}_{\rm eff}[{\mathcal P}]}
&=&{\mathcal Z}_0\left[\sqrt{\kappa}\frac{\beta(1+2v^2)}{\pi}\right]^{2N_l}\!
\!\!\int\! D[w_j] \; \exp
\left( -\frac{\beta}{2}\sum_{x, j}
w^\alpha_j(x)\, {\mathcal S}^{(2)}_{\alpha\beta}[{\mathcal P}] \, w^\beta_j(x)
\right.
\nonumber
\\
&& \hspace{30mm}
\left.
- \; \beta(1+2v^2)\sum_{x, j}
\Big[ \kappa^2 (w^4_j(x))^2 + (\vec{w}_j(x))^{2} \Big]
\right) \; ,
\la{def-5}
\end{eqnarray}
where $N_l=3N^3_sN_t$ is the number of links,  ${\mathcal
S}^{(2)}_{\alpha\beta}[{\mathcal P}] $ is a quadratic form from the Wilson action and we
used the equations of motion (\ref{eqm1}). We see that the result of integrating the
variables $\rho_j^\alpha(x)$ can be  formulated as a specific change of the Wilson
quadratic form.

It is convenient to write the complete quadratic form in the momentum  representation. We
expand the quantum fluctuations $w_j^\alpha(x)$ in plane waves:
\beq
w^\alpha_j(x)=\frac{1}{\sqrt{N^3_sN_t}}\sum_p e^{ip_4t+i \vec{p}\cdot \vec{x} } \;
\widetilde{w}^\alpha_j(p) \;\; , \;\;\;
p_4=\frac{2\pi k_4}{N_t} \;\; , \;\;\;   p_j=\frac{2\pi k_j}{N_s} \; ,
\eeq
where $k_4, k_j$ are integers: $k_4=0,1,\ldots N_t-1$, $k_j=0,1,\ldots N_s-1$.
The kernel  $W^{\alpha \beta}_{ij}(p)$ of the resulting quadratic form
has indices related to color ($\alpha, \beta=1,2,3,4$) and for the direction of the given
spatial link ($i,j=1,2,3$). For the trivial Polyakov line ($\varphi=0$) it is
diagonal in the color indices:
\beq
W^{\alpha \beta}_{ij}(p)=\delta^{\alpha\beta}\left[W^{(G)} \, \delta^{\alpha k}\delta^{\beta l}+
W^{(M)}_{ij}(p) \, \delta^{\alpha 4}\delta^{\beta 4}\right] \; ,
\eeq
($k,l=1,2,3$). The first term describes 3 massless degrees of freedom
which correspond to 3 gluons, the second one is related to a
massive excitation which is a lattice artifact,
\beq
W^{(G)}_{ij}(p)\!=\!(1\!-\!\cos p_4)\delta_{ij}+v^2\!\left[\delta_{ij} \Big[\!\cos p_{i}\!-
\!\sum_{l}\cos p_{l}\Big]\!-\!2(1\!-\!\delta_{ij})e^{i(p_{i}-p_{j})}\sin p_{i}\sin
p_{j}\right]+2v^2\delta_{ij} ,
\eeq
\beq
W^{(M)}_{ij}(p)\!=\!(1\!-\!\cos p_4)\delta_{ij}+v^2\!\left[\delta_{ij}\Big[\!\cos p_{i}\!-
\!\sum_{l}\cos p_{l}\Big]\!-\!2(1\!-\!\delta_{ij})e^{i(p_{i}-p_{j})}\cos p_{i}\cos
p_{j}\right]+2\kappa^2v^2\delta_{ij} .
\eeq
In these expressions the first term comes from the expansion of time-like plaquettes,
the second is due to the expansion of space-like plaquettes and the third is the
result of integrating over the $\rho_j^\alpha(x)$.
The integral (\ref{def-5}) turns into a Gaussian one,
\[
e^{-\beta{\cal S}_{\rm eff}[{\mathcal P}= \mathds{1}]}=
{\mathcal Z}_0\left[\sqrt{\kappa}\frac{\beta(1+2v^2)}{\pi}\right]^{2N_l} \int \!\! D[\widetilde{w}]
\exp\bigg( \! -\frac{\beta}{2}\sum_p\left[ \widetilde{w}^4_i(p)\, W^{(M)}_{ij}(p) \,
\widetilde{w}^4_j(-p)
\,
\right.
\]\beq
\left.
\hspace{30mm}
+ \; \vec{\widetilde{w}}_i(p) \, W^{(G)}_{ij}(p) \, \vec{\widetilde{w}}_j(-p)
\right]
\bigg) \; .
\la{def-6}
\eeq
The kernel $W^{(G)}$ of the quadratic form for the gluons has 3 eigenvalues,
\beq
\lambda_{1,2}=2-2\cos p_4+2v^2(3-\cos p_1-\cos p_2-\cos p_3), \qquad   \lambda_3=2-2\cos p_4 \; .
\eeq
The first two correspond to two transverse gluons, the third one to a longitudinal
gluon.  The eigenvalue of $W^{(M)}$ cannot be determined analytically, but it is seen to
correspond to a massive particle  (the eigenvalue does not vanish at $p_\mu=0$) with real
mass (the eigenvalue is positive for all $p$).

The longitudinal eigenvalue $\lambda_3$ vanishes for all momenta with $p_4=0$ and
therefore  the integral (\ref{def-6}) diverges. As always, zero modes of the quadratic
form correspond to some continuous  symmetry of the problem. In our case it is the
symmetry under gauge transformations  independent of time. Indeed, the solution
(\ref{s-p}) of the saddle point equations is not unique, since any function of the form
\beq
H_j(x) \; = \; h \, S(\vec{x}\,) \, S^\dagger(\vec{x}+ \widehat{j} \,)\; \; , \; \; \;
V_j(x) \; = \; v \, S(\vec{x}\,) \, S^\dagger(\vec{x}+ \widehat{j} \,) \; ,
\eeq
with time independent gauge matrices $S(\vec{x}\,)$, is also a solution of the saddle
point  equations with the same action. For this reason fluctuations $w_j^\alpha(x)$ in
the directions  corresponding to $S(\vec{x}\,)$ give rise to zero modes. Of course this
does not mean that  the complete integral is divergent, but the expansion in
$w(x)_j^\alpha$, which we use, breaks down and we have to modify it.

The eigenfunctions corresponding to the zero modes of the quadratic form for
$w_j^\alpha(x)$ are:
\beq
\eta_{j}^{(p)}(x) \; = \; \frac{\xi_{j}(p)}{\sqrt{N_{t}N_{s}^{3}}} \, e^{i\vec
{p}\cdot \vec{x}}
\; \; , \; \; \;
\xi_{j}(p) \; = \; \frac{\left\{  e^{ip_{1}}-1,e^{ip_{2}}-1,e^{ip_{3}}-1\right\}
}{\sqrt{4(\sin^{2}\frac{p_{1}}{2}+\sin^{2}\frac{p_{2}}{2}+\sin^{2}\frac{p_{3}}{2})}} \; .
\eeq
In the continuum limit $\xi_j(p)=\frac{i p_j}{|\vec{p}|}$ is the vector of the
longitudinal gluon polarization. Let us introduce unity (a la Faddeev-Popov):
\beq
1 =  \int\!\! D[S] D[w_j]\; J_{SU(2)}[V_{j}] \; \delta \Big( V_{j}(x)
- S(x) \Big[v \mathds{1} + \vec{w}_j(x) \cdot \vec{\sigma} \Big] \, S^\dagger(x + \widehat{j}\,)\Big) \,
\delta \left(  \sum_x
\vec{w}_j(x)  \eta_{j}^{(p)}(x)\right) .
\la{unity}
\eeq
Here $J_{SU(2)}[V_{i}]$ is the Jacobian for changing from variables
$V_j(x)$ to variables $S(\vec{x})$ and $w_j(x)$. The second $\delta$-function restricts the
integration over the $w_j(x)$ to be orthogonal to the zero modes of the quadratic form.
Direct calculation gives the following result:
\beq
\log J_{SU(2)}[V_{i}] \, = \, N^3_s\log(2\pi^2)+\frac{3}{2}\sum_{{\bf p}}
\log\left(4v^2N_{t}\left[\sin^2\frac{p_1}{2}+\sin^2\frac{p_2}{2}+\sin^2\frac{p_3}{2}\right]\right) \; .
\eeq
The first term is the volume of $SU(2)$ to which the measure $D[S]$ is normalized.

Using \eq{unity} in the integral (\ref{def-6}) we see that the integrand does not
depend on $S$ and hence the integral over $D[S]$ gives unity. Performing the integral
over the $w_j^\alpha(x)$ we arrive at
\[
-\beta{\cal S}_{\rm eff}[{\mathcal P}\!=\!{\bf 1}]=\log{\mathcal Z}_0+
3N_{s}^3N_{t}\log(4[1+2v^2]^{3/2})-\frac{3}{2}\log\det\widetilde{W}^{(M)}+
\log J_{SU(2)}+\frac{3}{2}N^3_s\log\frac{\beta}{2\pi}
\]\beq
-3\sum_{p}\log \left( 4 \sin^2\frac{p_4^2}{2}
+4 v^2\left[\sin^2\frac{p_1^2}{2}+\sin^2\frac{p_2^2}{2}+\sin^2\frac{p_3^2}{2}
\right]
\right)-\frac{3}{2}\sum_{p_4\neq 0}\log \left( 4\sin^2\frac{p_4^2}{2} \right) ,
\la{def-7}
\eeq
where we introduced the renormalized quadratic form
$\widetilde{W}^{(M)}=W^{(M)}/[(1+2v^2)\kappa^2]$.  The remaining contribution from
massive gluons (third term in (\ref{def-7}))  appears to be small in a wide region of
$\beta$ values. The second term in (\ref{def-7}) is the one loop correction to the
leading order vacuum energy  ($\log{\mathcal Z}_0$), the 4-th and the 5-th are the result
of fixing the  longitudinal zero modes, the 6-th is the contribution of transverse
gluons, and the 7-th from longitudinal gluons.

\section{Effective potential at 1-loop}

For $SU(2)$ we parameterize the constant temporal link variables by the matrix $U_4(x) =
e^{i\frac{\varphi}{N_t}\sigma_3}$ such that the Polyakov line is given by   ${\mathcal
P}=e^{i \varphi \sigma_3}$. A non-trivial Polyakov line ($\varphi \neq 0$) modifies the
time-like plaquettes and the modification is equivalent to a shift of the temporal
momenta $p_4$  entering the momentum space kernel,
\begin{eqnarray}
W^{\mu\nu}_{ij}(p)&=&\delta^{\mu\nu}\left[\delta^{\mu 1}W^{(G)}
\left(p_4+\frac{\varphi}{N_t},\vec{p}\right)
\right.
\nonumber
\\
&&
\left.
+ \; \delta^{\mu 2} W^{(G)}\left(p_4-\frac{\varphi}{N_t},\vec{p}\right)
+\delta^{\mu 3} W^{(G)}(p_4,\vec{p}\,)+\delta^{\mu 3}W^{(M)}(p_4,\vec{p}\,)
\right] \; .
\end{eqnarray}
The quadratic form for the massive particles remains unchanged.

We see that in the presence of a non-trivial Polyakov line the gluon quadratic form has
only one zero mode. The reason is clear: The Polyakov line explicitly breaks the residual
gauge invariance down to $U(1)$. Correspondingly, at $\varphi\neq 0$ we have to  fix only
one zero mode. For this purpose we introduce unity similar to \eq{unity}, but only for
the third color component of $w_j^\alpha(x)$ and change the integration in $D[S]$
from  $SU(2)$ to
$U(1)$. The effect of fixing the zero mode reduces for the statistical sum to:
\beq
\log J_{U(1)}+\frac{1}{2}N^3_s\log\frac{\beta}{2\pi} \; ,
\eeq
where
\beq
\log J_{U(1)}(V_{i})=N^3_s\log(2\pi)+\frac{1}{2}\sum_{\vec{p}}
\log\left(4v^2N_{t}\left[\sin^2\frac{p_1}{2}+\sin^2\frac{p_2}{2}+\sin^2\frac{p_3}{2}\right]\right) \; .
\eeq
Here $2\pi$ is the volume of $U(1)$. Other modifications of \eq{def-7} for the case
$\varphi\neq 0$ are: i) the contribution of transverse and longitudinal gluons
should be represented as a sum of 3 terms with $p_4$ and $p_4\pm\frac{\varphi}{N_t}$.
ii) for the longitudinal gluon terms with $p_4\pm\frac{\varphi}{N_t}$ one should include
also $p_4=0$ in the summation over momenta.

The summation over $p_4$ in the expression for the statistical sum can be performed
by means of the formulae derived in the Appendix. We obtain for the contribution of
the transverse gluons:
\beq
\log{\mathcal Z}^{\rm transverse}= -
\sum_{\vec{p}}\Big[ 2 \log \left( 4\sinh^2 N_t\alpha+4\sin^2\varphi \right) +
\log\left( 4 \sinh^2 N_t\alpha \right) \Big] \; ,
\la{trnsv}
\eeq
where
\beq
\sinh^2 \alpha=v^2\left[\sin^2\frac{p_1}{2}+\sin^2\frac{p_2}{2}+\sin^2\frac{p_3}{2}\right] \;.
\eeq
For the contribution of the longitudinal gluons (including the contribution of the
Jacobian) one finds:
\beq
\log{\mathcal Z}^{\rm longitudinal}=-\sum_{\vec{p}} \left[
\log(4\sin^2\varphi)+\frac{1}{2}\log\left(\frac{N_{t}}{2\pi\beta\cdot 4\sinh^2(N_t\alpha)}\right)
\right] \; .
\la{long-0}
\eeq
The complete expression for the statistical sum is obtained as a sum  of $\log{\mathcal
Z}^{\rm transverse}$,  $\log{\mathcal Z}^{\rm longitudinal}$ and the first 3 terms from
\eq{def-7}.

Lattices accessible for Monte Carlo simulations have become quite large in recent years.
For this reason
one can use the approximation $N_s\to\infty$ and change  the sums over the spatial
momenta to integrals. We introduce new continuous variables:
\[
q_j\, = \, \frac{2}{a}\sin\frac{p_j}{2} \; , \; \qquad  \sinh^2\alpha \, = \,
v^2\, \frac{ (\vec{q} \,)^{2} \, a^2}{4} \; .
\]
Then expression (\ref{trnsv}) becomes:
\beq
\frac{\log{\mathcal Z}^{\rm transverse}}{N^3_s}=-\prod_j\int_{-2/a}^{2/a}
\frac{a\; dq_j}{2\pi\sqrt{1-\frac{q_i^2 a^2}{4}}} \left[
\log \Big( 4\sinh^2 N_t\alpha+ 4 \sin^2\varphi \Big) +\log ( 4\sinh^2 N_t\alpha ) \right],
\la{integr}
\eeq
and analogously for $\log{\mathcal Z}^{\rm longitudinal}$, \eq{long-0}. These
expressions correspond to the continuum limit for spatial momenta. At $q\sim 1/a$ they are
regularized by our (cubic) lattice.

The continuum limit also requires that $N_t\to\infty$. Usually this condition is
fulfilled much worse in a typical lattice simulation than $N_s \rightarrow \infty$.
Nevertheless, let us calculate
the expansion of \eq{integr} in this limit. We rewrite $\log{\mathcal Z}^{\rm transverse}$
in the form
\begin{eqnarray}
\frac{\log{\mathcal Z}^{\rm transverse}}{N^3_s}
&=&-2\prod_i\int_{-2/a}^{2/a}\frac{a\; dq_i}{2\pi\sqrt{1-\frac{q_i^2 a^2}{4}}}
\Big[\log\left(1-2\cos2\varphi e^{-2\alpha N_{t}}+e^{-4\alpha N_{t}}\right)+
\nonumber
\\
&&
+\log\left(1+e^{-4\alpha N_{t}}\right)+3N_t\alpha\Big].
\la{SB}
\end{eqnarray}
For large $N_t$ the main contribution to the integrals over the first two terms in the
parenthesis comes from small $q_ia\ll 1$. Performing the integrals we obtain:
\beq
\frac{\log{\mathcal Z}^{\rm transverse}}{N^3_s N_t}=
-6c_1-\frac{\pi^2}{3N_t^4}\left[-\frac{1}{5}+
4\left(\frac{\varphi}{\pi}\right)^2\left(1-\frac{\varphi}{\pi}\right)^2 \right]
+\ldots \; ,
\eeq
where $c_1$ is the constant
\beq
c_1=\prod_i\int_{-2/a}^{2/a}\frac{a\; dq_i}{2\pi\sqrt{1-\frac{q_i^2 a^2}{4}}}\;
\alpha(q)\approx 0.90003 \; .
\eeq
Expression (\ref{SB}) is the celebrated perturbative potential
in $\varphi$ coming from transverse gluons, which is well-known
in the continuum \cite{GPY}
(for a lattice analogue of these calculations see \cite{W} and \cite{forkrand}).
This potential is defined only for $0<\varphi<\pi$ and outside this interval
should be continued according to periodicity. The constant in square brackets
is precisely the Stefan-Boltzmann energy of the transverse gluons.
Expression (\ref{SB}) is only an approximation of \eq{trnsv} for large $N_t$.
Nevertheless, it works rather well even for $N_t=2$.

An analogous calculation of the longitudinal gluons results in
\[
\frac{\log{\mathcal Z}^{\rm longitudinal}}{N^3_s }=
c_2-\half\log \left( \frac{N_tv^2}{2\pi\beta} \right) -\log(4\sin^2\varphi) \; ,
\]
\beq
c_2=\prod_i\int_{-2/a}^{2/a}\!\!\frac{a\; dq_i}{2\pi\sqrt{1-\frac{q_i^2 a^2}{4}}}
\log\left( \frac{4\sinh^2\alpha}{v^2}
\right) \; \approx \;  1.67339 \; .
\la{long-1}
\eeq

We are not so much interested in the statistical sum,
but in the free energy ${\cal F}$ given as
\beq
{\cal F} = -T \log{\mathcal Z} = \beta T {\cal S}_{\rm eff}=V_3\frac{1}{a^4}f \; .
\eeq
Being an extensive quantity, the free energy is proportional to the volume
$V_3 = (N_sa)^3$ of the system.  It is convenient to introduce the dimensionless
quantity $f$, which is a specific free energy in lattice units. Using
$T=1/(N_ta)$ it is simply given by
\beq
f \; = \; -\frac{\log{\mathcal Z}}{N^3_sN_t} \; .
\eeq
We are interested in this quantity as a function of $\varphi$ and temperature
$T/T_c=N_t^{(c)}/N_t$ in units of the critical temperature $T_c$ (here $N^{(c)}_t$ is the
value of $N_t$ at which the phase transition takes place). Collecting all terms we arrive at:
\beq
f=\varepsilon_{\rm vac} + \frac{\log\left(\frac{N_t}{2\pi\beta}\right)-2c_2}{2N_t}+
\frac{\log(4\sin^2\varphi)}{N_t}
+\frac{\pi^2}{3N_t^4}\left[-\frac{1}{5}+4\left(\frac{\varphi}{\pi}\right)^2
\left(1-\frac{\varphi}{\pi}\right)^2
\right].
\la{fren-1}
\eeq
Here $\varepsilon_{\rm vac}$ is the vacuum energy density in the 1-loop approximation:
\begin{eqnarray}
\varepsilon_{\rm vac}&=&3\beta\left[6+ v^2(1+3 v^2)
-\frac{1}{\beta}\log\left(\frac{2I_1(h)}{h}\right)\right]-
3\log\left(4(1+2v^2)^{3/2}\right)
\nonumber
\\
&&
+\;6c_1+\frac{3}{2N^3_sN_t}\log\det\widetilde{W}^m+\ldots
\la{vac1}
\end{eqnarray}
The first term here is the mean field approximation, \eq{mf},
and the subsequent terms represent the 1-loop
correction. As it was already said, the last term is
rather small and one can neglect it.

Other terms in \eq{fren-1} are corrections in $1/N_t$ and hence depend on the
temperature. We see that the effective potential on the lattice contains not only the
well-known perturbative potential \cite{W,forkrand} and the Stefan-Boltzmann energy, but
also additional terms coming from longitudinal gluons. These terms are linear in the
temperature (or depend as $T\log T$) and the coefficient in front of them is ultraviolet
divergent (it is proportional to $a^{-3}$). This contribution disguises the contribution
of transverse gluons to the energy and the effective potential, as at all $N_t$ (even for
$N_t=2$) it appears to be much larger than the perturbative potential and the
Stefan-Boltzmann energy.

The effective potential \eq{fren-1} diverges at $\varphi\to 0$. The reason is obvious: at
$\varphi\to 0$ we return to the situation where the symmetry of the saddle point field
restores back to $SU(2)$ and we have to fix 3 zero modes instead of one. At
$\varphi=0$ the statistical sum is given by \eq{def-7}. As compared to
\eq{fren-1} only the contribution of the longitudinal gluons changes
\beq
\frac{\log{\mathcal Z}^{\rm longitudinal}(\varphi=0)}{N^3_sN_t}=
\frac{3c_2}{2N_t}-\frac{3}{2N_t}\log\left(\frac{2\pi
N_t}{\beta}\right)+\frac{\log(2\pi^2)}{N_t} \; .
\la{long-2}
\eeq
In fact, it is clear that there is a transitional region at small $\varphi$ where neither
\eq{long-1} nor \eq{long-2} are applicable.

In this region one can still try to fix 3 modes (one zero mode and two with non-zero but
small eigenvalues) introducing unity according to \eq{unity}. However, the action at
non-zero $\varphi$ will depend on a $SU(2)$ matrix $S$. This dependence comes from
time-like plaquettes and leads to the following integrals over unitary matrices $S$  located
on spatial sites,
\beq
{\mathcal Z}_S(\varphi)=\int DS(\vec{x})\exp\left( \beta v^2 N_t
\sum_{\vec{x},{j}}\left[\half\Tr\left[ S^\dagger(\vec{x})e^{i\frac{\varphi\tau_3}{N_t}}
S(\vec{x})S^\dagger(\vec{x}+\widehat{j}) e^{-i\frac{\varphi\tau_3}{N_t}}
S(\vec{x}+\widehat{j})\right]-1
\right] \right) ,
\la{zs}
\eeq
(we here neglect the quantum fluctuations $w$). Another modification necessary for
the contribution of longitudinal gluons is that in the sum of logarithms of longitudinal
eigenvalues $\lambda_3$ the value $p_4=0$ should be omitted for all 3 colors. Finally also the
Jacobian
$J_{SU(2)}$ should be taken into account. We find
\beq
\frac{\log{\mathcal Z}^{\rm longitudinal}(\varphi)}{N^3_s}=
-\log\!\left(\frac{4\sin^2\varphi}{4\sin^2\frac{\varphi}{N_t}}\right)+\frac{1}{2}\log N_t+\frac{3}{2}\log\frac{\beta}{2\pi}+\frac{3}{2}c_2+\log2\pi^2+
\frac{\log{\mathcal Z}_S(\varphi)}{N^3_s}.
\la{long-3}
\eeq
At $\varphi=0$ when ${\mathcal Z}_S=1$ this expression reduces to \eq{long-2}. Let us pay
attention to the fact that it contains terms which are not periodic in $\varphi$ with
period $\pi$. This means that $\log{\mathcal Z}_S(\varphi)$ is also not periodic to
cancel this non-periodicity.

As already noted above,  the integrand in \eq{zs} does not depend on the matrices $S$ belonging to
the $U(1)$ subgroup of $SU(2)$: $S=e^{is_3\tau_3}$. At $\varphi\sim 1$  \eq{long-3} should reduce
to \eq{long-1}. Parameterizing an $SU(2)$ matrix as $S=e^{is_3\tau_3}e^{is_1\tau_1+is_2\tau_2}$ we
see that at $\varphi\sim 1$ the  integral over $s_1,s_2$ can be calculated by the saddle point
method. Expanding in $s_{1,2}$ we obtain
\footnote{The factor $\pi^{-1}$ in the measure is the ratio of $(2\pi)$,
which is volume of $U(1)$ and $2\pi^2$ which is the volume of $SU(2)$.}
\beq
{\mathcal Z}_S=\int\!\! \frac{ds_1(\vec{x})ds_2(\vec{x})}{\pi}
\exp\left(-2\beta
v^2\sin^2\frac{\varphi}{N_t}\sum_{\vec{x},{j}}\left[(s_1(\vec{x})-s_1(\vec{x}+\widehat{j}))^2+
(s_2(\vec{x})-s_2(\vec{x}+\widehat{j}))^2\right]\right).
\eeq
This integral can be easily calculated in the momentum representation
\beq
{\mathcal Z}_S=\exp\left( 8\beta v^2\sin^2\frac{\varphi}{N_t}
\sum_{\vec{p}}\Big[s_1(\vec{p}\,)s_1(-\vec{p}\,)+s_2(\vec{p}\,)s_2(-\vec{p}\,)\Big]
\Big[\sin^2\frac{p_1}{2}+\sin^2\frac{p_2}{2}+\sin^2\frac{p_3}{2}\Big]
\right).
\eeq
Integration results in:
\beq
\frac{{\mathcal Z}_S}{N^3_s}=
-c_2-\log4\sin^2\frac{\varphi}{N_t}-\log\frac{\beta}{2\pi}-\log\pi
\la{weak}
\eeq
This expression is also not periodic in $\varphi$ as expected. Substituting \eq{weak}
into \eq{long-3} we return to \eq{long-1} as it should be. However, we see that
\eq{long-1} is valid for not too small $\varphi$ where the coupling constant in the
Gaussian integration in $s_{1,2}$ is small enough.

To cover also the region of small $\varphi$,
let us change in the integral determining ${\cal Z}_S$
to the new variables:
\beq
{\mathcal Z}_S(\varphi)\!=\!\int\!\! D[\vec{n}]\;\delta(\vec{n}^2-1)
\exp\!\left(\!\half N_t\beta v^2\sin^2\frac{\varphi}{N_t}\sum_{\vec{x},{j}}
\left[\vec{n}(\vec{x})\cdot\vec{n}(\vec{x}+\hat{j})-1\right]\!\right)
,\;
n_a\!=\!\frac{1}{2}\Tr[S^\dagger\tau_3 S\tau_a].
\la{o3}
\eeq
Here $\vec{n}$ is a three-component vector with $\vec{n}^{\,2}=1$. The sum over
$j$ in this formula is running over all 6 directions $j = \pm 1, \pm 2, \pm3$.
Correspondingly, we divide the sum by a factor of two.
Integration over the unit vector $\vec{n}$ is an integration over
the factor-group $SU(2)/U(1)$.

Expression (\ref{o3}) is the statistical sum for a classical lattice theory --
the $O(3)$ $\sigma$-model. The role of the inverse coupling constant is played by:
\beq
\beta_{O(3)}=N_t\beta v^2 \sin^2\left(\frac{\varphi}{N_t}\right) \; \approx \;
\beta\frac{v^2\varphi^2}{N_t} \; .
\eeq
The approximation \eq{long-1} for the contribution of the longitudinal gluons corresponds to
the weak coupling limit of \eq{o3}.  We can also construct the strong coupling expansion for
the statistical sum:
\beq
\frac{\log{\mathcal Z}_S(\varphi)}{N^3_s}=-3\beta_{O(3)}+\frac{1}{2}\beta_{O(3)}^2+\ldots
\la{strong}
\eeq
This theory was the subject of many lattice studies (see, e.g. \cite{yot}) but we were not able
to find results for its free energy as a function of the coupling constant which are needed
for our purposes.

\begin{figure}
\begin{center}
\includegraphics[width= 9cm]{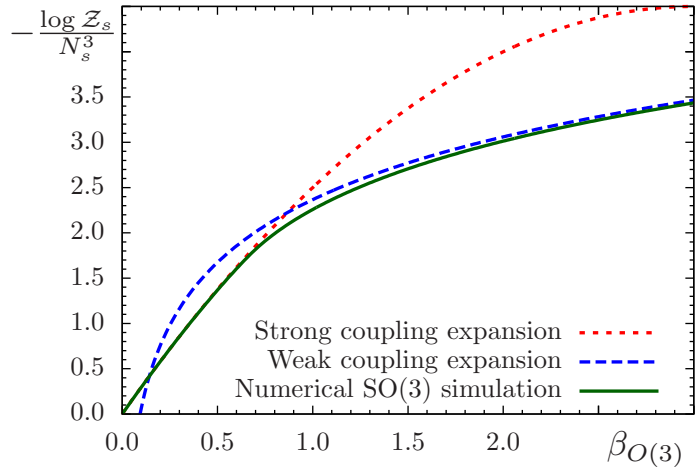}
\end{center}
\caption{$- \log{\cal Z}_S / N^3_s$ as a function of $\beta_{O(3)}$.
We compare the weak coupling expansion
(dashed blue curve), strong coupling expansion
(dotted red curve) and results of a numerical simulation on $64^3$ lattices
(full green curve).}
\label{fig:0}
\end{figure}

Still this calculation is a relatively easy task for lattice simulations
and we performed a corresponding simulation on $64^3$ lattices. In Fig.~\ref{fig:0}
we display the logarithm of the statistical sum for 3-dimensional $SO(3)$
theory together with the strong and weak coupling expansion results.
It appears that the weak coupling approximation works well in a wide
range of $\beta_{O(3)}$ values (see Fig.~\ref{fig:0}).

\section{Comparison with numerical simulations}

A numerical evaluation of the effective action $S_{\rm eff}({\cal P})$ for gauge groups
$SU(2)$ and $SU(3)$  was presented recently in \cite{DGS}. Relatively large lattices with
sizes up to $40^3\times N_t$ for $N_t=2,\ldots 20$ were used in a fixed scale approach
(i.e., fixed inverse gauge coupling $\beta$ and thus fixed lattice spacing $a$, with the
temperature driven by varying $N_t$). All temporal links, except those on the last time
slice, were fixed to be $U_4(x)= \mathds{1}$.  At the final time slice the temporal link
variables were suitably parameterized, using, e.g.,  $U_4(\vec{x},N_t-1) =
e^{i\varphi\sigma_3}$. In other words, the setup of \cite{DGS} precisely corresponds to
the one described in our Section II.
\begin{figure}[htb]
\begin{center}
\includegraphics[width=9 cm]{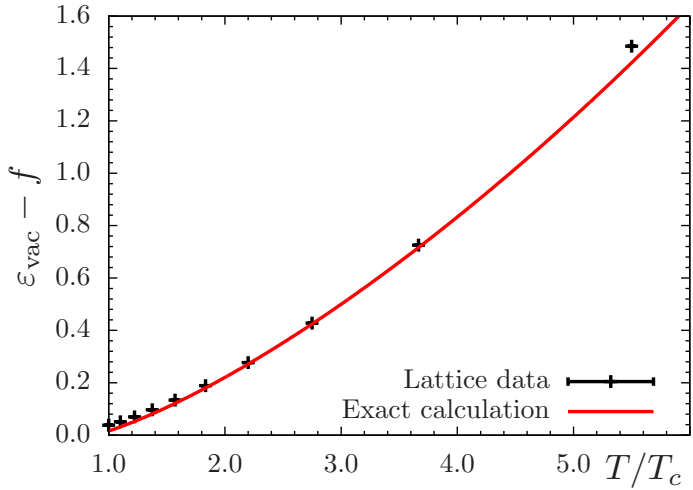}
\end{center}
\caption{Free energy for $\varphi=0$  as a function of temperature $T/T_c$ (vacuum energy
subtracted). The value of $\varepsilon_{vac}-f$ is plotted. The data points are from
\cite{DGS} and the full curve corresponds to \eq{fren-1}. }
\label{fig:1}
\end{figure}

The free energy of the theory was calculated by numerically integrating the averaged
plaquette expectation  value over $\beta$ starting from $\beta=0$ up to the value of
$\beta$ one wants to work at and which sets the lattice spacing $a$, i.e., the cutoff.
The corresponding values of the inverse coupling were $\beta=2.6$ for $SU(2)$  and
$\beta=6.2$ for $SU(3)$. Results were presented as a function of $\varphi$ and $T/T_c$. The
overall magnitude for the free energy was off by the aforementioned factors of 20 to  30
for the two gauge groups, but suitable ratios were in very good agreement with
perturbative results.

For $SU(2)$ the phase transition  was observed near $N_t=11$, so that the
maximal temperature studied with $N_t = 2$
was $T/T_c \sim 11/2 = 5.5$. Corrections from the discretization in the spatial
directions can easily be estimated and were
found \cite{DGS} to be very small for $N_s = 40$.

In order to describe data obtained in \cite{DGS} without forming ratios and to account
for the missing factors, we now apply the mean field
approximation developed above. This will be done only for the $SU(2)$ case.

\begin{center}
\begin{figure}[htb]
\begin{minipage}[t]{0.49\textwidth} 
\includegraphics[width=\textwidth]{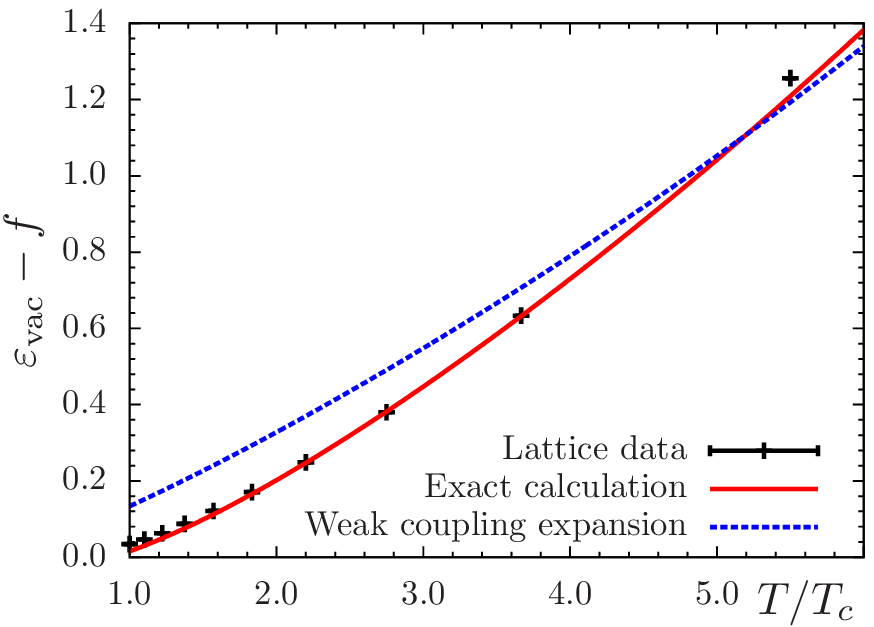}
\end{minipage}
\hspace{1.5mm}
\begin{minipage}[t]{0.49\textwidth} 
\includegraphics[width=\textwidth]{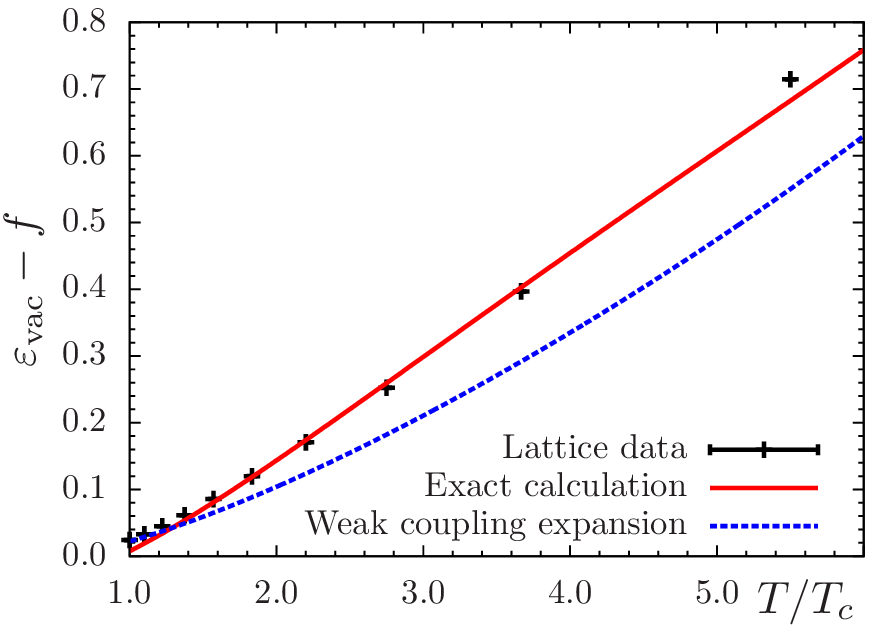}
\end{minipage}
\caption{Free energy for $\varphi=\frac{\pi}{8}$ (left) and $\varphi=\frac{\pi}{4}$ (right)
as a function of temperature $T/T_c$ (vacuum energy subtracted).
The data points are from \cite{DGS} and the full curve corresponds to \eq{fren-1}.}
\label{fig:2}
\end{figure}
\end{center}

To begin with, we compare the vacuum energy densities $\varepsilon_{\rm vac}$ to get
a first estimate
for the accuracy of
the mean field approach. Extrapolating the data to low temperatures we obtain:
\beq
\varepsilon^{(\rm exp)}_{\rm vac}=10.52 \; .
\la{vac3}
\eeq
At the same time, for $\beta=2.6$ \eq{vac1} gives
\beq
\varepsilon^{(\rm 0)}_{\rm vac}=12.51, \qquad \varepsilon^{(\rm 1)}_{\rm vac}=9.63 \; ,
\la{vac4}
\eeq
for the leading order and the 1-loop approximation. Based on these numbers we
estimate the accuracy of the mean field 1-loop approximation to be at the 10\% level.
\begin{center}
\begin{figure}[htb]
\begin{minipage}[t]{0.49\textwidth} 
\includegraphics[width=\textwidth]{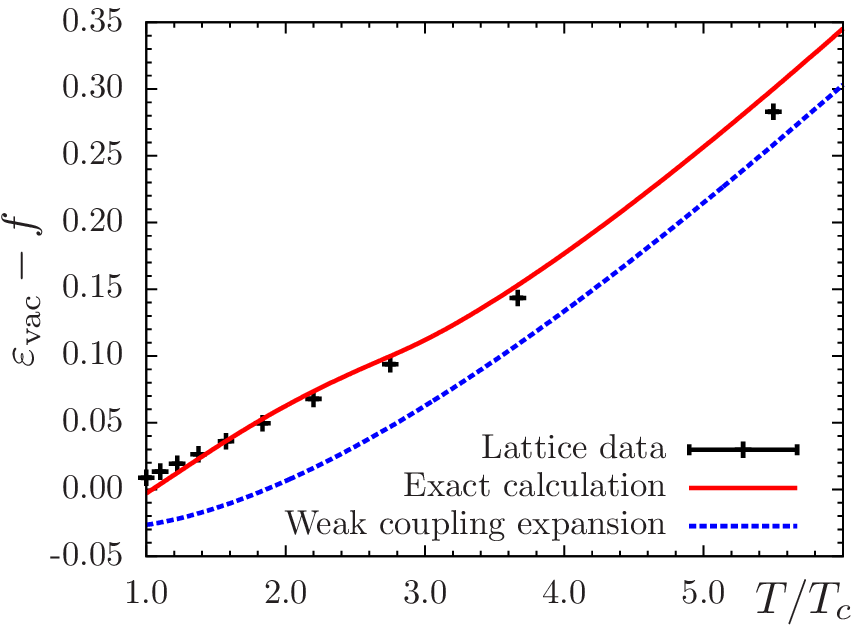}
\end{minipage}
\hspace{1.5mm}
\begin{minipage}[t]{0.49\textwidth} 
\includegraphics[width=\textwidth]{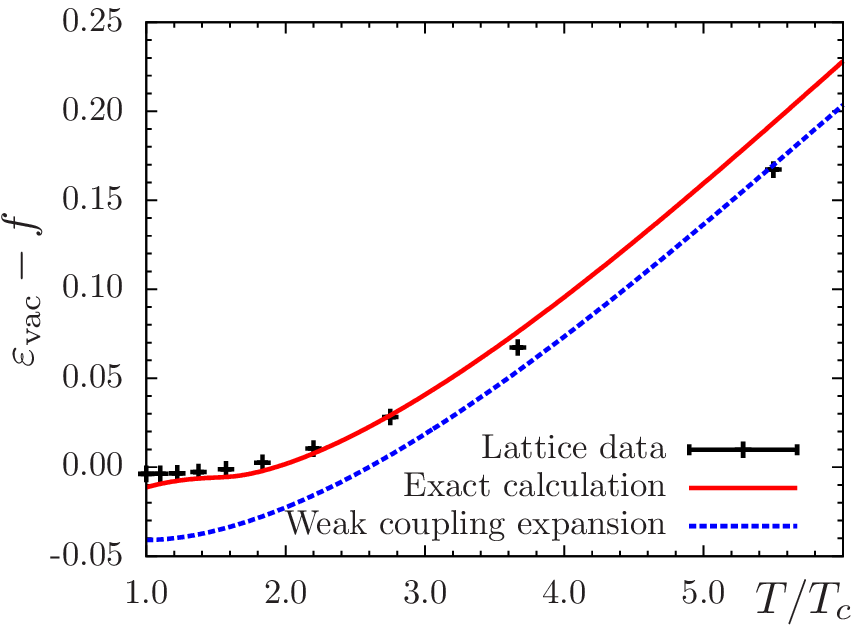}
\end{minipage}
\caption{Free energy  for $\varphi=\frac{3\pi}{8}$ (left)
and $\varphi=\frac{\pi}{2}$ (right) as a function of
temperature $T/T_c$ (vacuum energy subtracted).
The data points are from \cite{DGS} and the full curve corresponds to \eq{fren-1}.}
\label{fig:3}
\end{figure}
\end{center}

Next we inspect the temperature
dependence of the free energy at different values of $\varphi$.
In particular the values $\varphi=0,\frac{\pi}{8},\frac{\pi}{4},\frac{\pi}{2}$
were studied in \cite{DGS} and are compared to our calculations in
Figs.~\ref{fig:1}, \ref{fig:2} and \ref{fig:3}. In the plots we
subtract the vacuum energy and start at temperatures $T>T_c$,
since below $T_c$ mean field approximation cannot be expected to apply.

At $\varphi=0$ (see Fig.~\ref{fig:1}) the data are described rather well
by our results. The small discrepancy at
the largest $T$ may be understood from the fact that the corresponding $N_t=2$ is too small  for
using the expansion in $1/N_t$.

The results at $\varphi\neq 0$ are more ambiguous, as they should include the knowledge of the free
energy for the $O(3)$ sigma-model. Dotted blue curves represent the approximation \eq{fren-1}
which should be
valid at sufficiently large $\varphi$ and in red the answer corresponding
to the full $O(3)$ model in
$d=3$. We see that the numerical simulations have sufficient accuracy to distinguish these two
cases, though the weak coupling expansion is not bad for the whole range of temperatures and
$\varphi$.  The complete mean field theory works very well in all cases above the deconfinement
transition. The contribution of the perturbative potential (and the Stefan-Boltzmann energy) is
rather small on the scale used in the above pictures.

\section{Conclusions}

The effective potential for the Polyakov line ${\cal P}(\vec{x})$ is an interesting
quantity which easily can be measured on the lattice. A detailed analysis of its
properties should help with understanding quark confinement since the Polyakov line is
the order parameter for the deconfinement transition.

Unfortunately, as is well-known, the Polyakov line is ill-defined due to ultraviolet
divergences. In particular the Coulomb energy of the static charge, which is linearly
divergent, contributes to the Polyakov line.  Correspondingly, its average should be
exponentially suppressed for small lattice spacing in both, the confining and
deconfined phases. It is thus not surprising that the effective potential for such a
quantity is also ultraviolet divergent. In this paper we identified a corresponding
singularity originating in the contribution of longitudinal gluons.  Taking into account
this singularity, the effective potential has a very deep minimum at ${\cal P}(\vec{x}) =
\mathds{1}$ which always dominates. This ultraviolet divergence completely obscures the
perturbative potential even at very high temperatures and strongly distorts the
effective Lagrangian for the Polyakov line. It is this singularity which gives the main
contribution to the lattice simulations of \cite{DGS}, such that in \cite{DGS} only
suitable ratios of observables could be matched to perturbative results.

On the other hand it is clear that this problem is completely due to the poor definition
of the Polyakov line as an order parameter in  Yang-Mills theory. One has to introduce
some other quantity which is ultraviolet stable (see, e.g. \cite{renorm} and references therein)
and try to investigate the corresponding effective Lagrangian . The simplest way out is to subtract 
the mentioned UV divergence
(calculated according to the formulae of this paper) from the data and identify the
remaining  piece as the effective potential for the Polyakov line. However, it seems that
the current accuracy of the performed simulations  is not yet  sufficient for this
procedure.
 
\vskip 0.7true cm

\noindent
{\bf Acknowledgments}\\
V.P.\ acknowledges partial support by the grant RSGSS-4801.2012.2. H.-P.\ S.\ is supported by the FWF Doctoral Program
{\it Hadrons in Vacuum, Nuclei and Stars}, DK W-1203. The authors are grateful to
P.V.\ Pobylitsa for fruitful discussions.

\appendix
\section{Summation formulae}
In the text we need the summation of series of the form
\beq
\sum_{p} f(\cos p) \; ,
\la{eq1}
\eeq
where the sum is running over momenta $p_k=\frac{2\pi k}{N_t}$ with integers $k=0,1,\ldots N_t-1$.
To calculate such sums let us consider the integral,
\beq
\int_\Gamma\!\frac{dz}{2\pi i} \, \frac{ N_t \, z^{N_t-1}}{z^{N_t}-1}\,
f\left(\half[z+z^{-1}]\right) ,
\la{eq2}
\eeq
where the contour $\Gamma$ envelops the unit circle. The integrand in \eq{eq2} has poles
at $z=e^{i 2\pi k/N_t }$, and maybe additional ones from the function $f$. The residues
of the poles $z=e^{i 2\pi k/N_t }$ coincide with the terms of the sum \eq{eq1}. According
to Cauchy's theorem the integral can be evaluated in two ways:  Taking the residues
inside the contour $\Gamma$ (the sum \eq{eq1} and possibly singularities of $f$) and
outside the contour (singularities of $f$ only).  Comparison of these two ways of
computing the integral gives suitable summation formulas for \eq{eq1}.

By this method one can, e.g., obtain the general formula,
\beq
\sum_{p}\frac{1}{\sin^2\left(\frac{p}{2}+\frac{\varphi}{N_t}\right)+\sinh^2\alpha} \; = \; \frac{N_t}{\sinh 2\alpha}\;
\frac{\sinh 2N_t\alpha}{\sinh^2\alpha+\sin^2\varphi} \; .
\la{eq3}
\eeq
For $\varphi=0$ this formula (without derivation) was given in \cite{W}. Next, we
consider the sum,
\beq
\sum_{p}\log\left( 1+\gamma\sin^2\left(\frac{p}{2}+\frac{\varphi}{N_t}\right)\right) \; = \;
\int_0^\gamma\frac{d\gamma'}{\gamma'}\left[
N_t-\sum_p\frac{1}{1+\gamma'\sin^2\left(\frac{p}{2}+\frac{\varphi}{N_t}\right)}\right] \; .
\eeq
Using \eq{eq3} to calculate the sum and changing variables according to $(\gamma')^{-1}=\sinh^2\alpha'$
we obtain a calculable integral and arrive at:
\beq
\sum_p\log\left[\sin^2 \left(\frac{p}{2}+\frac{\varphi}{N_t} \right)+\sinh^2\alpha
\right] \; = \; \log\left(\frac{\sin^2\varphi+\sinh^2N_t\alpha}{2^{2N_t-2}}\right) .
\la{eq4}
\eeq
This is the main formula which we use to perform all summations in the text.
We use also particular cases of this formula:
\beq
\sum_{p\neq 0}\log\left(\sin^2 \left(\frac{p}{2}+\frac{\varphi}{N_t} \right)
\right ) = \log\left(\frac{\sin^2\varphi}{\sin^2\frac{\varphi}{N_t}}\right), \qquad
\sum_{p\neq 0}\log\sin^2\frac{p}{2}
= 2\log N_t \; .
\eeq

\end{document}